\documentclass[
journal=jpccck,
manuscript=article,
layout=twocolumn
]{achemso}

\usepackage[T1]{fontenc} 
\usepackage{graphicx} 
\usepackage[english]{babel}
\usepackage{bm}
\usepackage[usenames,dvipsnames]{xcolor}
\usepackage{hyperref}
\hypersetup{
    colorlinks,%
    citecolor=blue,%
    linkcolor=blue,%
    urlcolor=blue
}

\newcommand{\iPr}{$^\textrm{\scriptsize B}$NHC$^{\textrm{\scriptsize iPr}}$}
\newcommand{\Et}{$^\textrm{\scriptsize B}$NHC$^{\textrm{\scriptsize Et}}$}
\newcommand{\Me}{$^\textrm{\scriptsize B}$NHC$^{\textrm{\scriptsize Me}}$}
\newcommand{\X}{$^\textrm{\scriptsize B}$NHC$^{\textrm{\scriptsize X}}$}

\author{Felipe Crasto de Lima}
\email{felipe.lima@lnnano.cnpem.br}
\author{Adalberto Fazzio}
\email{adalberto.fazzio@lnnano.cnpem.br}
\affiliation{Brazilian Nanotechnology National Laboratory (LNNano), Brazilian Center for Research in Energy and Materials (CNPEM), 13083-970, Campinas, São Paulo, Brazil}
\author{Alastair B. McLean}
\affiliation{Department of Physics, Engineering Physics and Astronomy, Queen's University, Kingston, ON, K7L3N6, Canada.}
\email{mcleana@queensu.ca}
\author{Roberto H. Miwa}
\affiliation{Instituto de F\'isica, Universidade Federal de Uberl\^andia, \\ C.P. 593, 38400-902, Uberl\^andia, MG,  Brazil}
\email{hiroki@ufu.br}

\title{Simulation of XANES spectroscopy and the calculation of total energies for N-heterocyclic carbenes on Au(111)}
\abbreviations{IR,NMR,UV}

\begin{document}

\begin{tocentry}

\includegraphics[scale=1]{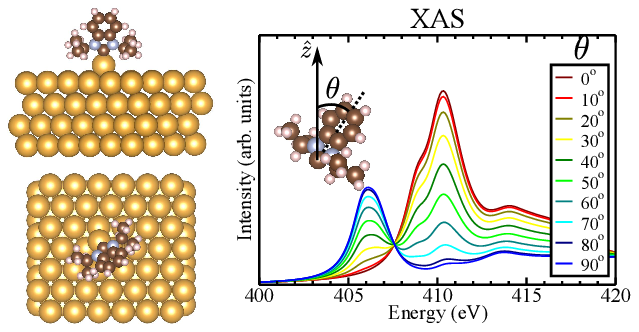}

%
%
%
 
\end{tocentry}

\begin{abstract}
It has recently been demonstrated that N-heterocyclic carbenes (NHCs) form self-assembled monolayers (SAMs) on metal surfaces. Consequently, it is important to both characterize and understand their binding modes to fully exploit NHCs in functional surface systems. To assist with this effort, we have performed {\it first-principles} total energy calculations for NHCs on Au(111) and simulations of X-ray absorption near edge structure (XANES). The NHCs we have considered are N,N-dimethyl-, N,N-diethyl-, N,N-diisopropylbenzimidazolylidene (\X,  with X=Me, Et, and iPr, respectively) and the bis-\X\, complexes with Au derived from these molecules. We present a comprehensive analysis of the energetic stability of both the \X\,  and the complexes on Au(111) and, for the former, examine the role of the wing group in determining the attachment geometry. Further structural characterization is performed by calculating the nitrogen K-edge X-ray absorption spectra.  Our simulated XANES results give insight into (i) the relationship between the \X/Au geometry and the N($1s$) $\rightarrow$ $\pi^\ast/\sigma^\ast$, pre-edge/near-edge, absorption intensities, and (ii) the contributions of the molecular deformation and molecule-surface electronic interaction to the XANES spectrum. Our simulations are compared with recent experimental results. 
\end{abstract}

\section{Introduction}

There is currently interest in using N-heterocyclic carbenes (NHCs) to create self-assembled monolayers (SAMs) on metal surfaces \cite{WeidnerAJC2011, ZhukhovitskiyJACS2013, NATURECrudden2014, ZhukhovitskiyChemRev2015, NATURECrudden2016, RuhlingAC2016, NATUREWang2016, EngelChemSocRev2017, SalorinneAC2017, CPCLarrea2017, ManJACS2018}. NHC-SAMs have been shown to be more thermally and chemically stable than thiols \cite{NATURECrudden2014, NATURECrudden2016}, the ligand class of choice for the creation of SAMs \cite{HakkinenNature2012}. The utility of NHC-SAMs stems from the fact that NHCs have a divalent carbon atom with a lone pair; they are $\sigma$-donor ligands that form strong bonds to transition metals \cite{ZhukhovitskiyJACS2013}. Such stronger bond can lead to an increased thermal robustness of current SAMs application, for instance in organic electronics\cite{CSRcasalini2017} electrochemical sensing \cite{JSSEMandler2011}, and catalysis \cite{ACRschoenbaum2014}. Moreover, by changing the N,N-substituents, or wing groups, their properties can be readily tuned, allowing for the rational design of 2D structures and properties\cite{CRSmith2019, NATCOMMliang2019}.

A number of experimental studies have been performed with {\iPr}. For instance, in the {\iPr}/Au(111) system the {\iPr}-Au bond strength was estimated, using thermal desorption spectroscopy (TDS), to be (158 $\pm$ 10) kJ mol$^{-1}$ \cite{NATURECrudden2016}. The first images of NHC self-assembly were obtained with {\iPr} on Au(111) \cite{NATURECrudden2014} and high resolution electron energy loss spectroscopy (HREELS) was used to demonstrate that the attachment geometry is upright \cite{NATURECrudden2016} when it is deposited on Au(111) at room temperature in ultra-high vacuum (UHV). Consequently, it is likely that {\iPr} adopts an atop geometry, bonding to a surface Au atom, or a geometry where it bonds to an adatom on the Au(111) surface \cite{wangNatChem2017,CPCLarrea2017}. The latter geometry places the wing groups further from the surface and, therefore, reduces the steric interaction between the molecule and the surface. However, in a recent study, Lovat {\it et al.} also considered tilted and flat-lying \iPr geometries \cite{CSLovat2019}. This approach finds support from recent experimental studies that have accumulated evidences for different binding modalities of  NHC on metal surfaces\cite{CRSmith2019}. Meanwhile, in a very recent study, that employed NHCs with different combinations of side groups and a range of molecular deposition/annealing  protocols, Inayeh {\it et al.}\cite{CHEMRXIVinayeh2020} found that only \iPr\, presents structurally stable  dual conformation on the Au(111) surface; namely attached to the surface Au adatom, or lying flat on the Au(111) surface.  Here, the atomic scale understanding of  the binding modes in NHC/surface systems, in particular  the role played by the side groups, will be helpful not only to  support the experimental findings, but also to provide insights that will facilitate the design of new molecular self-assemblies on metal surfaces.

\begin{figure}[h!]
\includegraphics[width=\columnwidth]{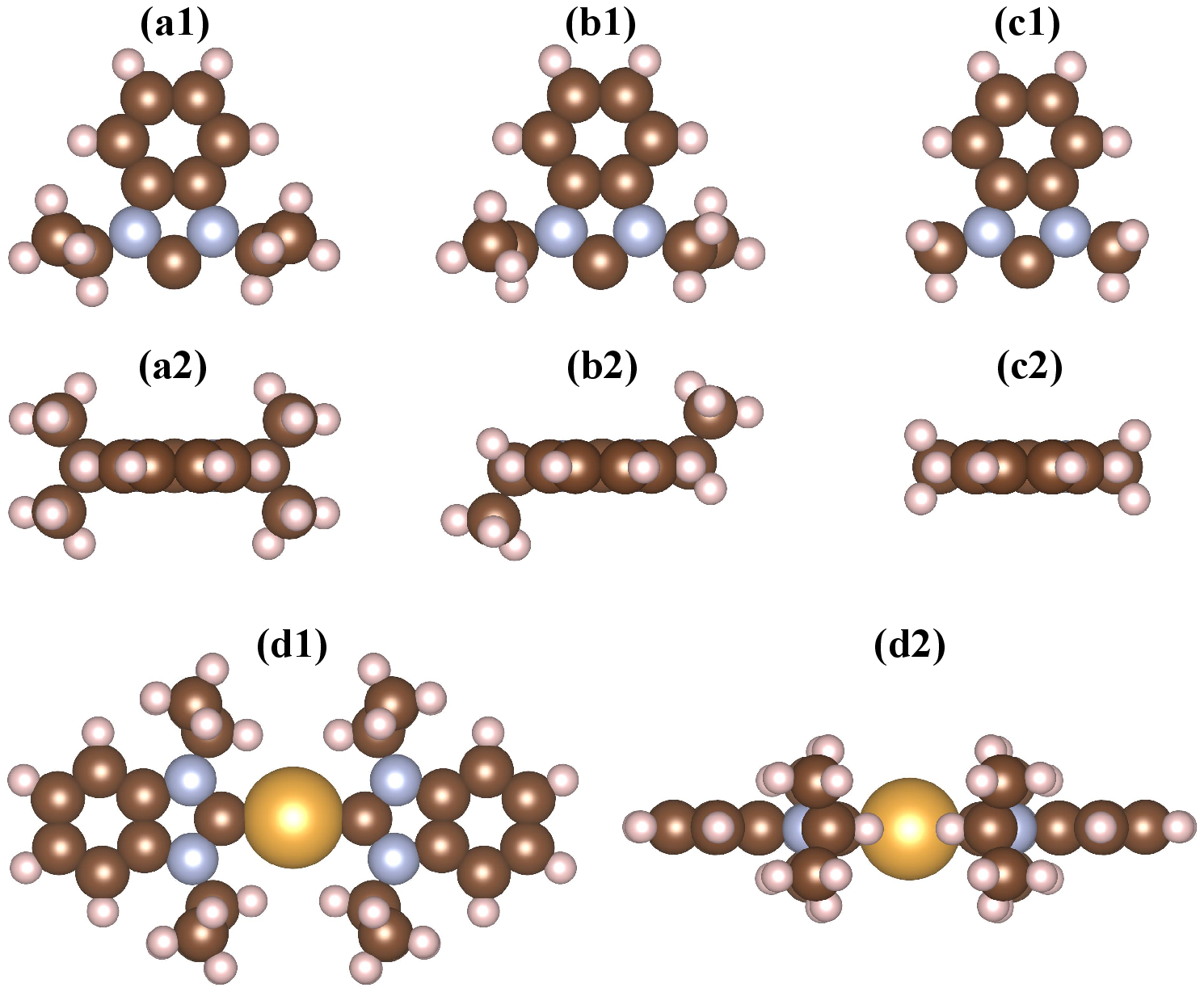}
\caption{\label{mol-str} 
The atomic structure of the NHCs investigated: (a) {\iPr}, (b) {\Et}, and (c) {\Me}. (a1)--(c1) View parallel (a2)--(c2) perpendicular to the molecular plane. Panels (d1) and (d2) show the {\iPr}-Au-{\iPr} complex.}
\end{figure}

Spectroscopy techniques have been used in order to examine  the electronic interactions, and structural properties of  molecules adsorbed on solid surfaces. For instance, (i) X-ray photoelectron spectroscopy experiments  addressing the electronic charge transfers and the characterization of molecule-surface chemical bonds of NHCs on metal surfaces \cite{jiangChemSci2017,krzykawskaACSNano2020}; and (ii) near-edge X-ray adsorption fine-structure spectroscopy (NEXAFS), recently used to characterize the conformation of NHCs on Au(111) \cite{CSLovat2019}. However it is clear that having simulated XANES spectra for candidate binding geometries, combined with total energy calculations, would be particularly helpful as an aid to data interpretation. Consequently, in this paper we have performed {\it ab-initio} calculations based on Density Functional Theory (DFT) in order to understand the binding modes of different benzimidazol based NHCs, namely \X{} where X = \{iPr,  Et, Me\}, and bis-NHC complexes \X-Au-\X\, (Fig.\,\ref{mol-str}), on the Au(111) surface.  We show a  correlation of the NHC's wing groups and its binding mode; in particular for \iPr\, we present a detailed total energy picture that supports the experimental finding of dual conformation on Au(111).  Further structural characterization was performed through simulations of    nitrogen K-edge X-ray near edge structure (XANES) spectra. The relationship between the molecular conformation and the  pre-edge/near-edge absorption spectra were interpreted in terms of the  projection of  the density of states, while the role played by the molecule-surface electronic interactions was examined through  XANES simulations of  hypothetical molecular configurations.

\section{Methods}

The calculations of atomic geometry were performed based on the DFT approach, as implemented in the VASP code \cite{VASP}. The exchange correlation term was described using the GGA functional proposed by Perdew, Burke, and Ernzerhof \cite{PBE}. The Kohn-Sham orbitals are expanded in a plane wave basis set with an energy cutoff of $400$\,eV. The Brillouin zone is sampled according to the Monkhorst-Pack method \cite{PhysRevB.13.5188}, using a gamma-centered $3 \times 3 \times 1$ mesh. The electron-ion interactions are taken into account using the Projector Augmented Wave method \cite{PAW}. We have considered an Au(111) slab of 4-layers in a $5 \times 5$ supercell with the two lower layers fixed in the bulk lattice parameter. All NHC/Au geometries have been relaxed until atomic forces were lower than 0.02 eV/{\AA}, with van der Waals interactions included within the vdW-DF approach \cite{PhysRevLett.92.246401}. Although the optB88-vdW functional \cite{klimesJPhysC2010} is more suited to describe the benzene binding energy onto the Au surface \cite{JCPcarrasco2014}, we found that this functional together with vdW-DF2 \cite{VDW-DF2} overestimate the NHC/Au binding energy by $\approx 60 \%$ \cite{CSLovat2019}, while the vdW-DF method is in excellent agreement with experimental results \cite{NATURECrudden2016}.

The Nitrogen K-edge X-ray absorption spectroscopy (XANES) was calculated using the XSPECTRA package \cite{PhysRevB.66.195107}, within the Quantum ESPRESSO code \cite{qe}, by using the Gauge-Including Projector Augmented-Wave (GIPAW) method\cite{PRBpickard2001} to calculate the dipolar K-edge\cite{JPCMbrouder1990},
\begin{equation}
\sigma (\omega) \propto \omega \sum_n |\langle n | \hat{\bm e} \cdot {\bm r} | {\rm 1s} \rangle|^2 \delta (\varepsilon_n - \varepsilon_{\rm 1s} - \hbar \omega),
\end{equation}
where $|n \rangle$ and $|{\rm 1s} \rangle$ are the final and 1s orbital wavefunction with energies $\varepsilon_n$ and $\varepsilon_{\rm 1s}$, respectively. Such approaches allow to take into account the core-hole effect of the 1s orbital of the X-ray absorber atom, while to obtain the X-ray absorption spectroscopy intensities unaffected by the presence of the pseudopotential\cite{PRBtaillefumier2002, PRBgougoussis2009} by reconstructing the all-electron wave function. In order to eliminate spurious interactions we have considered a distance of $14.7$\,{\AA} between core-holes and its periodic images.

\section{Results}

\begin{figure}[t]
\centering
\includegraphics[width=\columnwidth]{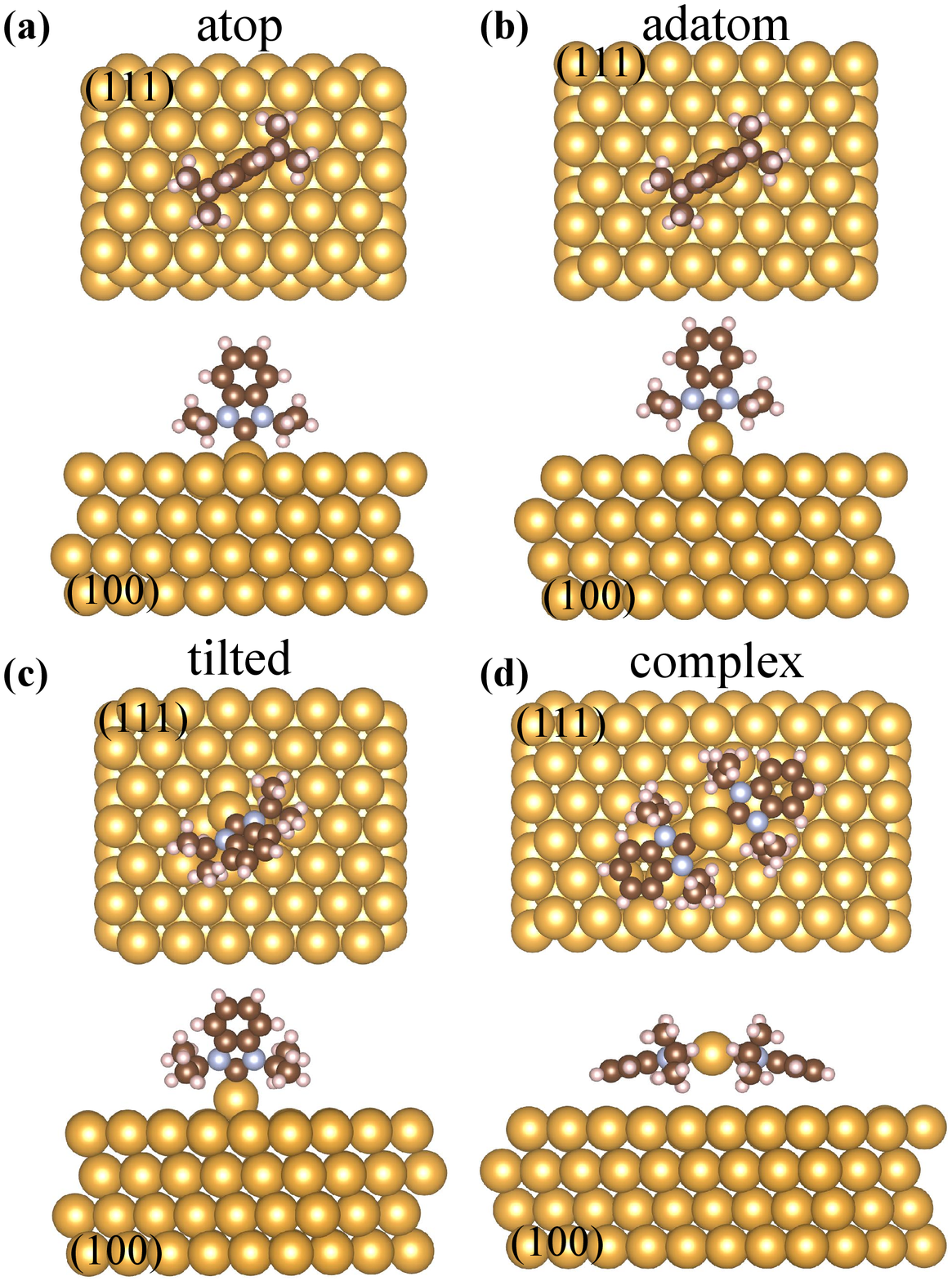}
\caption{\label{str} Atomic geometries for \iPr on Au(111), namely (a) atop-{\iPr}/Au, (b) adatom-{\iPr}/Au, (c) tilted-{\iPr}/Au and (d) complex-{\iPr}/Au. Here the top panels shown the Au(111) surface while the bottom panels show the Au(100) direction. These same configurations was considered for the {\Et} and {\Me} cases.}
\end{figure}

\paragraph{Total Energy Calculations.} We have considered four different adsorption geometries of \X\, on the Au(111) surface (\X/Au, with X=iPr, Et, and Me), as shown in   Figs.~\ref{str}(a)-(d). In \emph{atop} and \emph{adatom}, the molecule is aligned with the surface normal, and bonded to a surface Au atom (\emph{atop}) or to a surface adatom (\emph{adatom}), Figs.~\ref{str}(a) and (b) respectively. In \emph{tilted} [Fig.~\ref{str}(c)], the molecule is bonded to a surface adatom but rotated towards the surface and in \emph{complex} two molecules form a bis-NHC complex {\X}--Au--{\X} [Figs.\,\ref{mol-str}(d1)-(d2)] that lies flat or nearly flat on the Au(111) surface [Fig.~\ref{str}(d)]. 

The binding energy ($E^b$) is a measure of the stability of the {\X}/Au adsorption geometry, where, as usual, $E^b$ is defined to be the total energy difference between {\X}/Au and the separated components, 
\begin{equation}
E^b = E({\rm ^BNHC^X})+E(\textrm{Au}) - E({\rm^B NHC^X/\textrm{Au}}). 
\end{equation}

\begin{table}[h]
\centering
\caption{\label{tab1} {\X} Binding energy $E^b$ (eV/molecule) to Au(111) surface, molecule electron loss $\Delta q$ (e), deformation energy $E^\textnormal{def}$ (meV/molecule), carbon-gold distance $d$ ({\AA}). The molecule--surface vertical distance $h$ ({\AA}) for \emph{complex}-{\X}/Au is given in parenthesis.}
\begin{tabular}{lccc}
\hline
 & \multicolumn{3}{c}{{\iPr/Au}} \\
\cline{2-4}
System & E$^{b}$ & $d$ & $\Delta q$  \\
\hline
atop                  & 1.728 & 2.133 & 0.275  \\
adatom                & 2.600 & 2.058 & 0.296 \\
tilted                & 2.631  & 2.063 & 0.335  \\
complex (4.30\,{\AA}) & 2.578 & 2.070  & 0.229  \\
\hline
 & \multicolumn{3}{c}{{\Et}/Au} \\
\cline{2-4}
System          & E$^{b}$ & $d$ & $\Delta q$  \\
\hline 
atop                  & 1.534 & 2.127  & 0.285\\
adatom                & 2.501 & 2.055  & 0.332 \\
tilted                & 2.499 & 2.060  & 0.338  \\
complex (3.92\,{\AA}) & 2.636 & 2.059  & 0.229 \\ 
\hline
 & \multicolumn{3}{c}{{\Me}/Au} \\
\cline{2-4}
System & E$^{b}$ & $d$ & $\Delta q$ \\
\hline
atop                  & 1.310 & 2.131 & 0.331 \\  
adatom                & 2.265 & 2.051 & 0.298 \\  
tilted                & 2.282 & 2.057 & 0.351 \\  
complex (3.65\,{\AA}) & 2.532 & 2.062 & 0.234  \\  
\hline
\end{tabular}
\end{table}

Our calculations of the binding energy, summarized in Table~\ref{tab1}, reveal that the least stable geometry is \emph{atop} for all three NHCs, where  we found $E^b$ between 1.3 and 1.7\,eV, and the C--Au bond length ($d$) of $\approx$ 2.13\,{\AA}. These results are in good agreement with ones presented in the current literature. For instance, the binding energy and the C--Au bond length of  1.30\,eV and 2.12\,\AA\, for \emph{atop}--\iPr/Au predicted by first-principles DFT calculations\,\cite{NATURECrudden2014}, and the subsequent experimental measurement of   $E^b$\,=\,1.64\,eV\,\cite{NATURECrudden2016}. Moreover, recent theoretical studies predicted $E^b$ of 1.47\,\cite{CAEJRodriguez2016}/1.25\,\cite{CMTang2017}\,eV and $d$=2.10/2.11\,\AA\, by including the long-range molecule--surface dispersion interactions  using the semi-empirical DFT-D2 and -D3 approaches\,\cite{grimme2006semiempirical,grimme2010consistent}. The binding energy increases by  about 0.9\,eV in the \emph{adatom} and \emph{tilted} configurations, which is in agreement with the more reactive character of the Au adatom sites. At the equilibrium geometry,   the vertical position of the molecule increases  by about 1.3\,\AA\, with respect to the surface, reducing the  steric constraints at the \X/Au interface, and thus enabling the search for energetically more stable configurations. The \X\, molecule sits on the Au adatom, with a C--Au bond length of  2.06\,\AA, that in turn lies on the hollow-fcc site of Au(111), giving rise to the so called "ballbot-type" structure\,\cite{wangNatChem2017}. Our total energy results show that the \emph{adatom} and \emph{tilted} configurations present practically the same binding energies. Here, the search for the tilted geometry was performed by considering an initial tilt angle ($\theta_0$) of 30$^\circ$, and then the molecule and surface atomic  coordinates were fully relaxed as described above (Methods). We found, for the three \X/Au systems,  lowest energy configurations for tilt-angles $\theta$ of about 27$^\circ$ with respect to the surface normal; whereas,  by considering larger values for the initial tilt angle, for instance $\theta_0$\,=\,50$^\circ$, we found a  metastable configuration, by 0.226\,eV  for $\theta$\,$\approx$\,54$^\circ$.

\begin{figure}[h]
\centering
\includegraphics[width=\columnwidth]{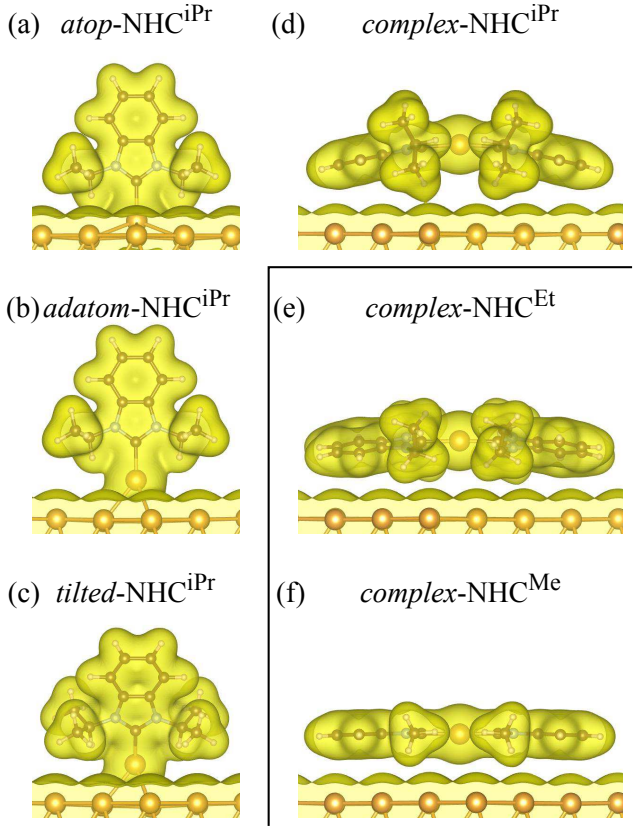}
\caption{\label{tot-charge} Total charge densities for (a) \emph{atop}--, (b) \emph{adatom}-- and (c) \emph{tilted}--{\iPr}/Au; and for \emph{complex}--{\X}/Au with X= (d) iPr, (e) Et and (f) Me. The isosurface scale is set to $0.01$\,e/\AA$^3$.}
\end{figure}

It is worth noting that, in addition to the formation of C--Au chemical bonds, the long range vdW interaction between the surface Au atoms and the H atoms, of the CH$_3$ side groups\,\cite{CMTang2017,JACSBakker2018}, also contributes to the energetic stability of the upstanding or nearly upstanding  \X/Au. As shown in Table\,\ref{tab1}, the binding energy increases proportionally with the size of the side wings, for example, we found  $E^b$\,=\,2.27, 2.50, and 2.60\,eV  for \emph{adatom}--\X/Au, with X=Me, Et, and iPr. The same occurs for the \emph{atop} and \emph{tilted} configurations.

In Figs.\,\ref{tot-charge}(a)--(c) we present the total charge density at the molecule-surface interface of the \emph{atop}--, \emph{adatom}--, and \emph{tilted}--\iPr/Au configurations.  Similar electronic distributions have been observed for the other molecules, where  we can identify the formation of C--Au chemical bonds, and  the emergence of an orbital  overlap between the wing-groups and the Au surface atoms  in  \emph{atop}--\iPr\, [Fig.\,\ref{tot-charge}(a)]. Such an  orbital overlap allow us to infer a  strengthening of the molecule-surface repulsive interaction, reflecting in a larger C--Au equilibrium bond length (2.13\,\AA) compared with those of  \emph{adatom} and \emph{tilted} geometries (2.06\,\AA).

As shown in Figs.\,\ref{tot-charge}(d)--(f), there are no charge density overlap at molecule-surface interface region of \emph{complex}--\X/Au. At the equilibrium geometry, we found  that the  horizontal  buckling  and molecule--surface distance ($h$) are proportional to the "size" of the wing groups, Table\,\ref{tab1}. The energetic stability   is dictated by  attractive long range  vdW interactions between the \X\, $\pi$ orbitals and  the  H atoms of the CH$_3$ side groups with the Au(111) surface. Our results of binding energy reveal that the wing groups rules the energetic preference for the \emph{complex}--\X\, geometry, namely  $E^b$ increases by 0.14 and 0.25\,eV/\X, for X = Et and Me, when compared with the one of the  \emph{tilted} geometry,  whereas it reduces by 0.05\,eV for \iPr. Thus, we can infer that that the formation of \emph{complex} is quite likely for \Et/ and \Me/Au whereas for \iPr/Au there is an energetic preference (although relatively small) for the \emph{tilted} configuration. These results support the recent experimental work, performed by Larrea {\it et al.}\,\cite{CPCLarrea2017}, addressing the orientation and the self-assembly of those molecules on Au(111). Similar results  on the role played by the  "size" of the NHC wing groups, on the molecular conformation  on Au(111), have been reported in a recent study  by Lovat\,\cite{CSLovat2019} {\it et al.}. In particular,  they suggested a thermally activated process in order to form \emph{complex} structures of \iPr. Indeed, in a very recent study, Inayeh {\it et al.}\cite{CHEMRXIVinayeh2020} have shown that, in addition to the  molecular  structure of the side groups, other physical factors such as the  surface coverage, annealing protocol, and molecular mobility also influence the conformation of \X\, molecules on Au(111). In particular, for X=iPr,   they found both \emph{tilted}  and \emph{complex} geometries, where the flat lying \emph{complex} structure was found to have a higher surface mobility that the "ballbot" structure of \emph{adatom}-- and \emph{tilted}--\iPr.

\begin{figure}[h]
\includegraphics[width=\columnwidth]{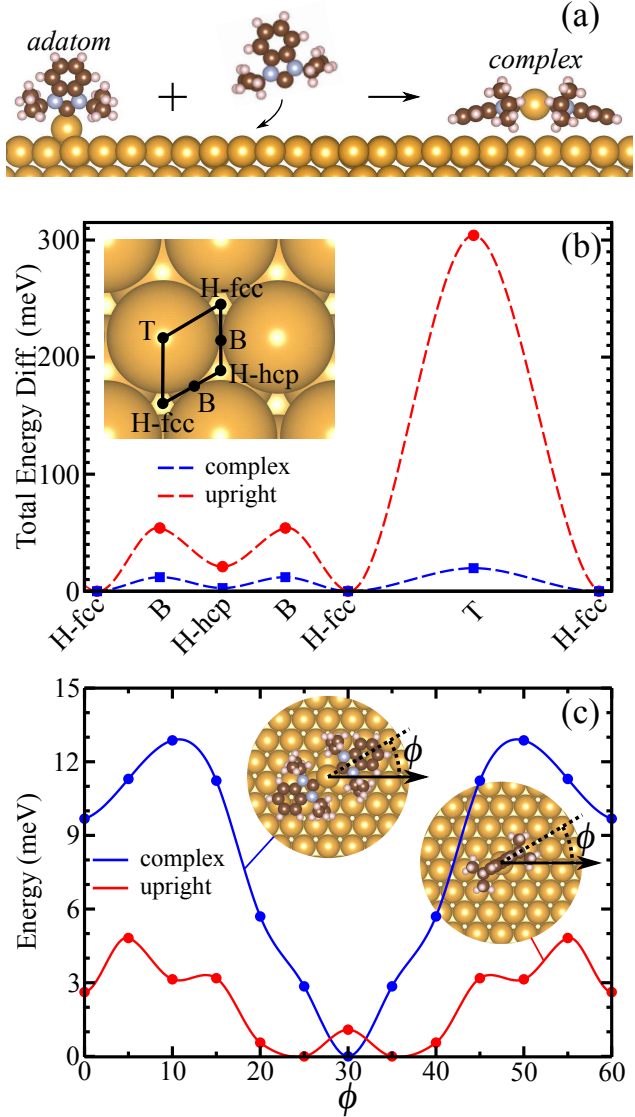}
\caption{\label{iPr} Complex formation picture (a); and energy barrier estimation for {\iPr} in different sites (b) and throughout an azimuthal rotation degree of freedom (c).}
\end{figure}

Such a dual behavior  of \iPr\, can be examined  by considering the schematic scenario presented in Fig.\,\ref{iPr}(a), combined with our binding energy results presented in Table\,\ref{tab1}.  That is, given an \emph{adatom} structure and a neighboring additional molecule, we may have the following final \iPr/Au structures, (i) the additional molecule attaches to the surface Au atom resulting in the \emph{atop} configuration, (ii) the formation of a second \emph{adatom} structure, if there were any source of additional Au adatoms, or (iii) results in a \emph{complex} structure.  Our calculated binding energies  indicate that in (ii) we have the largest energy gain, however, in the absence of additional (and accessible) Au adatoms we find that (iii) is more likely than (i) by 0.40\,eV/\iPr. Indeed,  similar reaction pathway has been proposed by Bakker {\it et al.}\,\cite{bakkerJACS2018} for \emph{complex}--NHC$^{\footnotesize\textrm{IMe}}$/Au.

In Fig.\,\ref{iPr}(b)  we present the total energy differences of \iPr\, on hole-fcc (H-fcc), -hcp (H-hcp), bridge (B), and top (T) sites. Due to the symmetry of these surface sites, our total energy results (although based on static calculations) allow us to infer the activation energy (AE) of upstanding \iPr\, on Au(111)\,\cite{wangNatChem2017}. Here, the molecular ballbot like diffusion of \emph{adatom}--\iPr\,  will take place passing through bridge sites facing AEs of  54 and 33\,meV for  the H-fcc\,$\rightarrow$\,H-hcp and H-hcp\,$\rightarrow$\,H-fcc diffusion paths; while the  T site is quite  unlikely for the  \emph{adatom} or \emph{tilted} configurations. In contrast, \emph{complex}--\iPr\, will face  nearly flat energy barriers, with AEs lower than 20\,meV, which is comparable with the ones involved with its rotation normal to  Au(111) surface, up to 14\,meV as depicted in Fig.\,\ref{iPr}(c). Such  lower  energy barriers confirm the experimentally observed nearly free diffusion of \emph{complex}--NHC$^{\footnotesize\textrm{IMe}}$\,\cite{jiangChemSci2017,wangNatChem2017} and --\iPr\, \cite{CHEMRXIVinayeh2020} on the Au(111) surface.   In the same diagram [Fig.\,\ref{iPr}(c)], for sake of comparison,  we present the rotation energy barrier for the upstanding \emph{adatom} configuration, which is lower than 6\,meV, indicating that the rotational degree of freedom plays a minor role on the calculated AEs. 
 
\begin{figure}[h]
\includegraphics[width=\columnwidth]{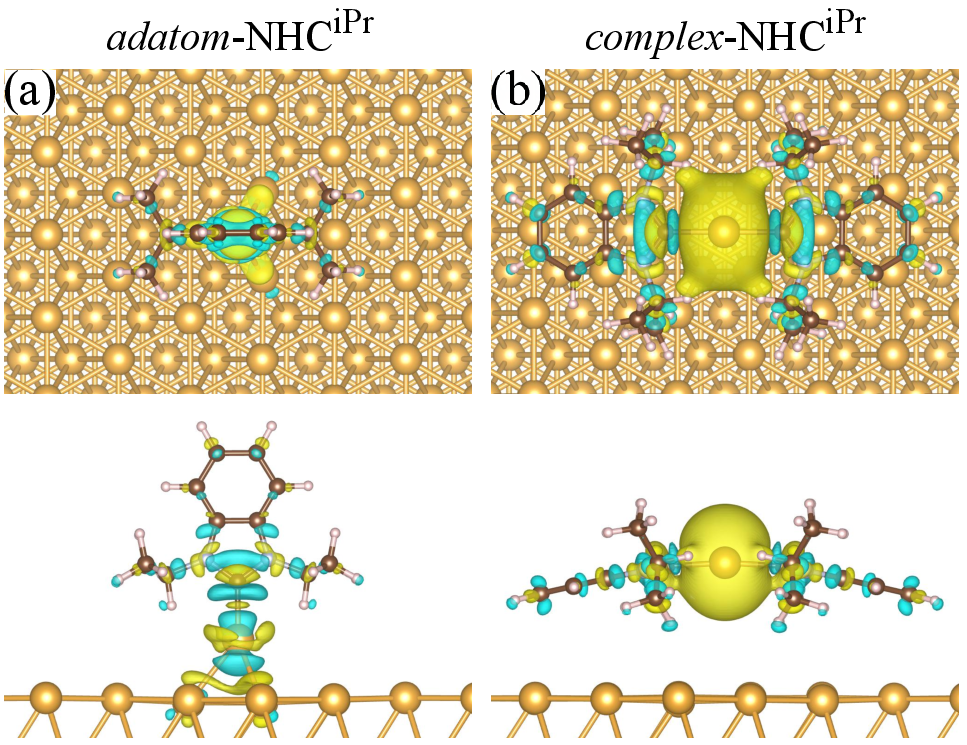}
\caption{\label{charge} Map of the total charge transfers for (a) \emph{adatom}--{\iPr}/Au and (b) \emph{complex}--{\iPr}/Au. The isosurface scale is set to $0.001$\,e/\AA$^3$.}
\end{figure}

\paragraph{Simulations of XANES.} In Figs.\,\ref{charge}(a) and (b) we present a map of the total charge transfers in the \emph{adatom}-- and \emph{complex}--\iPr/Au structures. By using the Bader charge density analysis, we find a net charge transfer of 0.30 and 0.23\,$e$ ($\Delta q$ in Table\,\ref{tab1}) from the \iPr\, molecule to the Au(111) surface. For both configurations, the total charge displacements take place mostly at the \iPr/Au interface region, spreading out not only  along  the \iPr\, backbone but also   in the  CH$_3$ side groups.  Similar results were obtained for the other \X/Au systems.  Such a molecule\,$\rightarrow$\,surface charge donation is in agreement with the  recent X-ray photo-electron spectroscopy (XPS) results, further supported by DFT calculations\,\cite{jiangChemSci2017,changJPCA2017,krzykawskaACSNano2020}. Indeed, each \X/Au system presents  its own spectroscopic fingerprint, characterizing the  molecular conformation and molecule-surface electronic interaction. For instance,  XANES measurements have been done addressing the molecular conformation of NHC$^\textrm{\footnotesize X}$ (X=dipp and Me) and \iPr\, on Au(111)\,\cite{CSLovat2019}.  Here, in order to provide a more complete picture of the  \X/Au  systems,  based on first-principles DFT simulations, we performed a detailed study of the nitrogen K-edge XANES spectra.

\begin{figure}[h]
\includegraphics[width=\columnwidth]{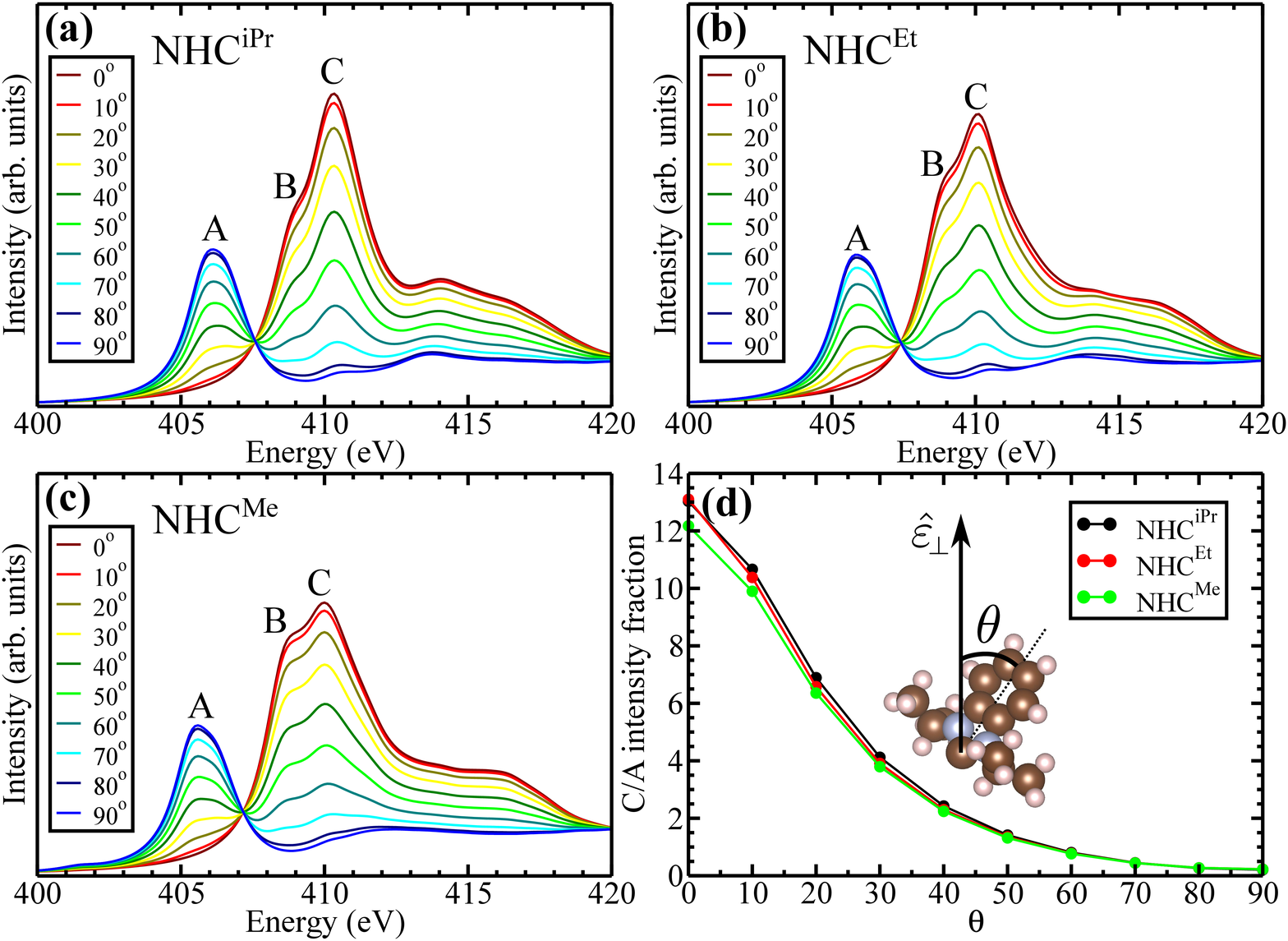}
\caption{\label{xas-theta} Polarization angle dependence of XANES for freestanding (a) {\iPr}, (b) {\Et} and (c) {\Me} molecules. (d) Fraction between C and A peaks as a function of the tilted angle. The angle is defined in relation to the benzene ring plane as shown in (d) inset. The energy scale is set in relation to the core-level energy.}
\end{figure}

First, we examine the absorption  spectra of (isolated) free-standing molecules, focusing on the relationship between the absorption spectra and the orientation of the molecule with respect to the polarization of the radiation ($\hat{\varepsilon}_{\perp}$), indicated by $\theta$ [inset  in Fig.~\ref{xas-theta}(d)]. In Figs.~\ref{xas-theta}(a)-(c) we present the XANES results for {\iPr}, {\Et}, and {\Me}, respectively. The K-edge spectra of the nitrogen atoms exhibits two main features, one associated with the N($1s$)\,$\rightarrow$\,$\pi^\ast$ transition, absorption peak A, and another associated with the N($1s$)\,$\rightarrow$\,$\sigma^\ast$ transition, giving rise to features B and C. The energy of feature A, which comprises a single peak  between 405 and 406\,eV,  weakly depends on the atomic structure of the wing groups, and its intensity reduces for lower values of $\theta$. That is, for $\theta=90^\circ$\,$\rightarrow$\,$0$  the N($1s$)\,$\rightarrow$\,$\sigma^\ast$ transitions start to dominate the XANES spectra. The intensities of B and C (between 408 and 411\,eV) increase, and we notice that these absorption features  are  sensitive to the  atomic structure of the wing groups attached to the nitrogen atoms. In Fig.~\ref{xas-theta}(d) we present the intensity ratio  ($\eta$) between the absorption peaks C (I$_{\rm C}$) and A (I$_{\rm A}$), $\eta={\rm I_C/I_A}$, as a function of the molecule orientation, $\theta$. We found that for $\theta$ varying from $0^\circ$ to $90^\circ$, the ${\rm I_C/I_A}$ absorption ratio reduces  from $\sim$12.5 to  0.2; we will show below that such a ratio $\eta$ can be used to estimate the orientation of the molecule with respect to the radiation polarization axis, $\hat{\varepsilon}_{\perp}$. 

\begin{figure}[h]
\includegraphics[width=\columnwidth]{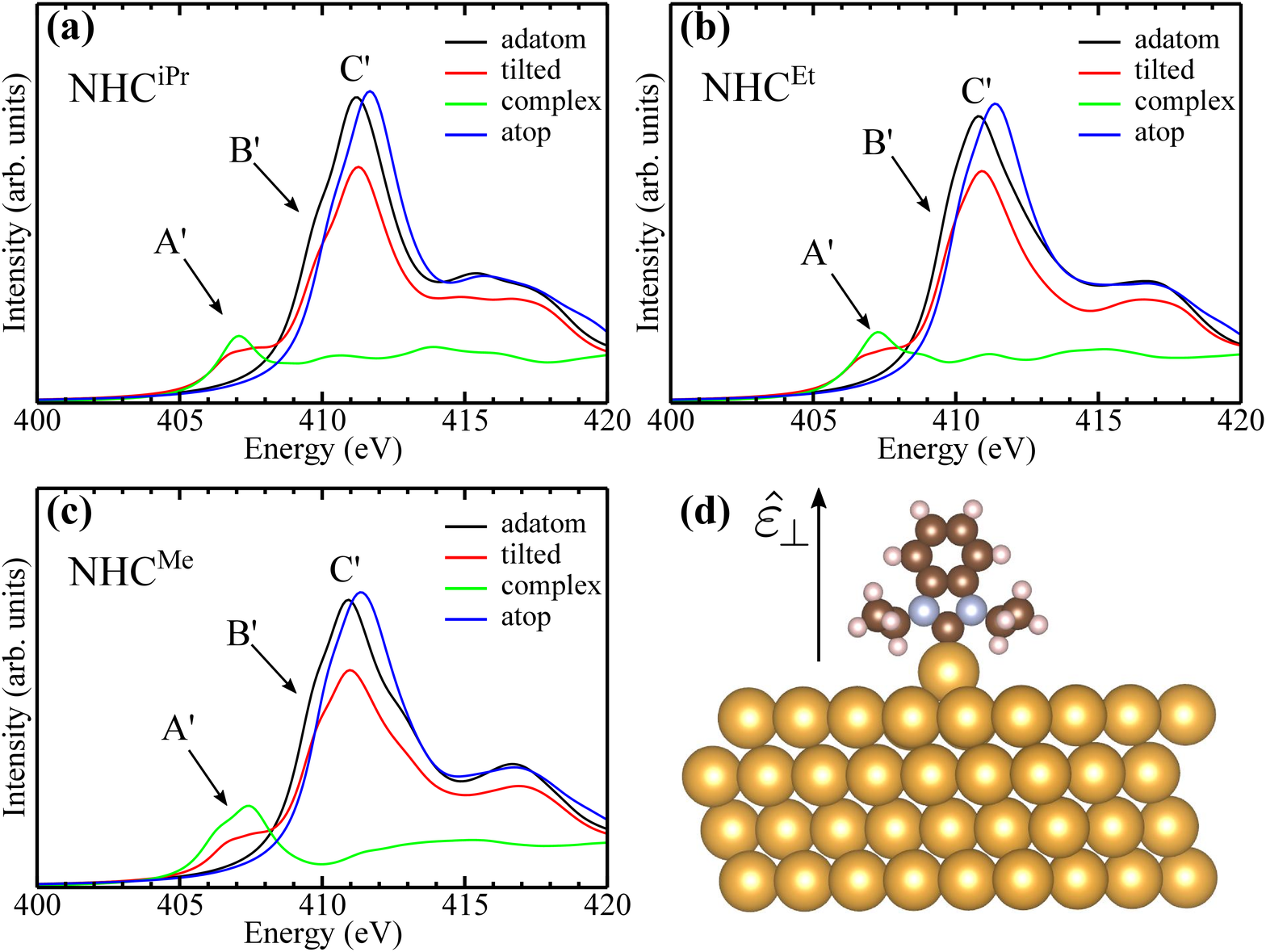}
\caption{\label{xas-new} XANES spectra for the configurations \emph{atop}-, \emph{adatom}-, \emph{tilted}- and \emph{complex}-{\X}/Au, with (a) X=iPr, (b) X=Et, and (c) X=Me. (d) XANES polarization vector perpendicular to Au(111) surface. The energy scale is set in relation to the core-level energy.}
\end{figure}

\begin{figure}[h!]
\includegraphics[width=\columnwidth]{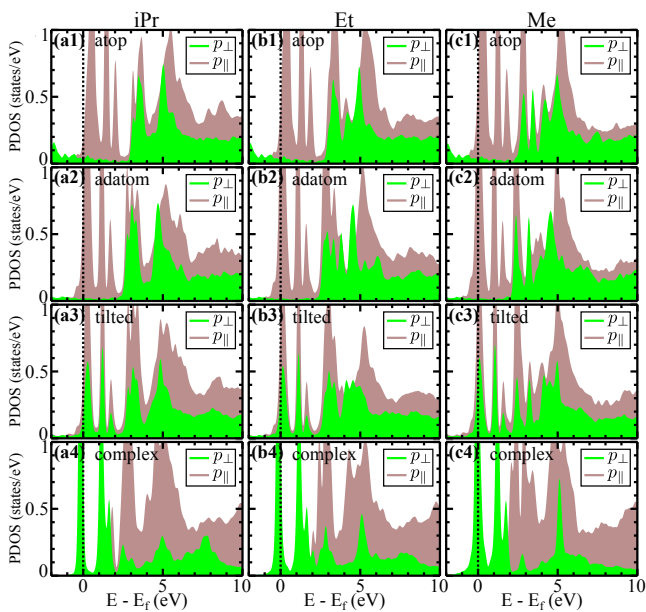}
\caption{\label{pdos} Projected density of states in the $p$ orbitals of N and neighboring C atoms. The $p_{\perp}$ ($p_{\parallel}$) are the $p$ orbitals perpendicular (parallel) to the Au(111) surface. The (a)-(c) panels are the PDOS for {\X}, and X=iPr, Et and Me, respectively. (a1)-(c1) Show the \emph{atop}-, (a2)-(c2) the \emph{adatom}-, (a3)--(c3) the \emph{tilted}-, and (a4)-(c4) the \emph{complex}-{\X}/Au configurations.}
\end{figure}

The features A, B and C of the free-standing molecules are in general preserved in \X/Au. In Fig.\,\ref{xas-new} we can identify their counterparts A$^\prime$, B$^\prime$ and C$^\prime$ for a polarization vector  perpendicular to the surface plane [$\hat\varepsilon_\perp$ in Fig.~\ref{xas-new}(d)]. The near-edge region of the absorption spectra is dominated by the features B$^\prime$ and C$^\prime$, while A$^\prime$ lies in the pre-edge region [Figs.~\ref{xas-new}(a)--(c)]. The  interpretation of these XANES spectra can be done through the calculation of the orbital projected density of states (PDOS) of the 2$p$ orbitals of  nitrogen,  and its nearest neighbor carbon atoms. In  Fig.\,\ref{pdos} we present our PDOS results, where  we have projected the $2p$ orbitals in two components, one  perpendicular ($p_{\perp}$) and another parallel ($p_{\parallel}$) to the Au(111) surface. 

For the upstanding configurations,  \emph{atop}-- and \emph{adatom}--{\X}/Au, the $\sigma^\ast$ ($\pi^\ast$) orbitals are projected on the  $p_{\perp}$  ($p_{\parallel}$) component, {\it i.e.} $\sigma^\ast$\,$\rightarrow$\,$p_{\perp}$ ($\pi^\ast$\,$\rightarrow$\,$p_{\parallel}$),  thus for $\hat\varepsilon=\hat\varepsilon_{\perp}$ the K-edge absorption  spectra will probe the N($1s$)\,$\rightarrow$\,$\sigma^\ast$ transitions.  As shown in Figs.~\ref{pdos}(a1)-(c1) and \ref{pdos}(a2)-(c2),  the electronic states near the Fermi level ($E_F$) project on the N and C $\pi^\ast$  orbitals, thus  consistent  with the absence of the pre-edge feature A$^\prime$, while the near-edge spectra B$^\prime$ and C$^\prime$ [Figs.\,\ref{xas-new}(a)-(c)] characterize the $\sigma^\ast$ orbitals which  start to rise at about $(E_F+2.5)$\,eV.  Unfortunately, the nitrogen K-edge spectra do not allow to distinguish the upstanding \emph{atop} and \emph{adatom} geometries,  unless the energy difference of $\sim$1\,eV within  the near-edge region, which can attributed to the C--N  bond (length/angle) distortions of the  side wings. In contrast, the \emph{tilted} configuration can be identified by the  rise of the feature A$^\prime$, concomitantly with the reduction of the absorption intensities of B$^\prime$ and C$^\prime$ when compared with those of the upstanding, \emph{atom} and \emph{adatom}, structures. We find that  A$^\prime$ comes from the N($1s$)\,$\rightarrow$\,$\pi^\ast$ transition, since the PDOS of the \emph{tilted}--\X/Au system [Fig.~\ref{pdos}(a3)-(c3)] show that  both $\sigma^\ast$ and $\pi^\ast$ orbitals  are projected on $p_{\perp}$, $\sigma^\ast + \pi^\ast$\,$\rightarrow$\,$p_{\perp}$. In the same PDOS diagram, it is noticeable the reduction of the $p_\perp$ component above $E_F+2.5$\,eV, compared with the ones  of the upstanding geometries, Figs.\,\ref{pdos}(a1)-(c1) and (a2)-(c2). We found that the  increase (reduction) of the $\pi^\ast$ ($\sigma^\ast$) orbital projection on the $p_\perp$ component becomes more evident in \emph{complex}--\X/Au [Figs.\,\ref{pdos}(a4)-(c4)]. Such  PDOS features reflect on the XANES spectra [Fig.\,\ref{xas-new}], here characterized by  an enhancement of the   A$^\prime$ absorption spectrum, which correspond to the N($1s$)\,$\rightarrow$\,$\pi^\ast$ transition to the $2p_z$  orbitals, of the N and nearest neighbor C atoms, lying at the Fermi level, followed by  a nearly suppression of the pre-edge absorption peaks, B$^\prime$ and C$^\prime$.

As discussed above, the \emph{tilted}-{\X}/Au structures are energetically more stable for $\theta$ about $27^\circ$ with respect to the surface normal direction.  As we have done for the free standing molecules, here we evaluate the intensity ratio between the absorption peaks C$^\prime$ and A$^\prime$, $\eta^\prime={\rm I_{C^\prime}/I_{A^\prime}}$. We obtained $\eta^\prime$\,=\,4.53, 4.50, and 4.11 for X=iPr, Et and Me. Interestingly, based on the $\eta={\rm I_{C}/I_{A}}$ ratio for free-standing molecules, Fig.~\ref{xas-theta}(d), those results of  $\eta^\prime$ correspond to tilt angles of about 30$^\circ$, 27$^\circ$, and 28$^\circ$, thus  nicely consistent with the calculated equilibrium geometries for the \emph{tilted}-\X/Au configurations. Thus, suggesting that  this ratio can serve as a guide to probe the angle of the molecule relative to the Au(111) surface.

\begin{figure}[t]
\includegraphics[width=\columnwidth]{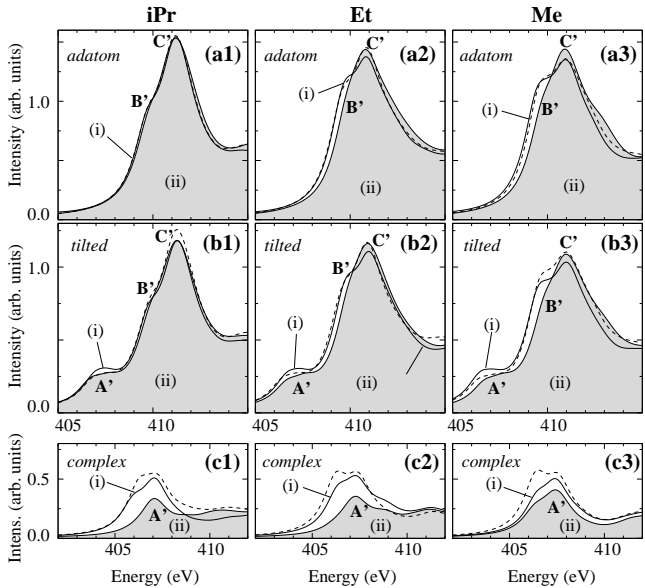}
\caption{\label{xas-ads-isol}  XANES spectra with (shaded curve) and without (continuous black lines) the Au substrate presence keeping the molecule/substrate atomic configuration and the isolated relaxed molecule (dashed lines). The iPr, Et and Me molecules are shown in (a1)-(a3), (b1)-(b3) and (c1)-(c3) respectively, while the \emph{adatom}, \emph{tilted} and \emph{complex} configurations in (a1)-(c1), (a2)-(c2) and (a3)-(c3) 
respectively. Here for a better comparison of each feature the spectra have been shifted in order to the peak A' coincide with the {\X}/Au configuration.}
\end{figure}

We can improve the understanding of the absorption spectra through  XANES simulations of   hypothetical molecular structures. Here, we have considered  an isolated \X\,  molecule constrained to the geometry of the adsorbed system. It   represents an imaginary configuration where the surface potential is "turned-off", but the molecule is not allowed to relax. The XANES  spectra of this system is indicated as (i) in Fig.\,\ref{xas-ads-isol}, and the shaded region  [(ii)] represents the absorption spectra of the final system, \X/Au.  Therefore, the comparison between (i) and (ii) reveals the role played by the molecule-surface electronic interaction in the nitrogen K-edge XANES spectra, such as the formation of C--Au chemical bonds, and \X\,$\rightarrow$\,Au(111) charge transfers. These comparisons for the upstanding and nearly upstanding  configurations are shown in Figs.~\ref{xas-ads-isol}(a) and (b).  We find that the effect of the molecule-surface interactions on the XANES spectra is negligible in \emph{adatom}--\iPr/Au [Fig.\,\ref{xas-ads-isol}(a1)], while there is a small reduction on the  absorption spectrum A$^\prime$ for the \emph{tilted} structure [Fig.\,\ref{xas-ads-isol}(b1)]. Similar reduction on the pre-edge  spectrum has been observed for the other molecules, Figs.\,\ref{xas-ads-isol}(b2) and (b3). Meanwhile, the near edge spectra of \emph{adatom}-- and \emph{tilted}--\X/Au, with X=Et and Me, are characterized by slight changes on the absorption peaks B$^\prime$ and C$^\prime$, the absorption intensity of the former (latter) reduces (increases). Further XANES simulations of the   fully relaxed free-standing molecules, indicated by dashed lines in Fig.\,\ref{xas-ads-isol}, show that the molecular relaxations are characterized by very small contributions to the final \X/Au absorption spectra of the upstanding or nearly upstanding \X/Au systems. In contrast, in  \emph{complex}--\X/Au, where the molecule lies flat or nearly flat on the Au(111) surface, we find that the molecule-surface interactions [(i)\,$\rightarrow$\,(ii)], and the change of the equilibrium geometry of the molecule [dashed-line\,$\rightarrow$\,solid-line] play  important roles on the final XANES spectra. As shown in Fig.\,\ref{xas-ads-isol}(c1)-(c3), both processes reduce the intensity of the absorption feature A$^\prime$. These greater (lower) sensibility   of the pre-edge (near-edge) absorption features A$^\prime$ (B$^\prime$ and C$^\prime$) reflects the larger (smaller)  molecule-surface interaction between the molecular $\pi$ ($\sigma$)  orbitals, of the nitrogen and the nearest neighbor carbon atoms, and the 6$s$ orbitals of the Au surface atoms. These results indicate that the use of $\eta^\prime(={\rm I_{C^\prime}/I_{A^\prime}})$ in order to infer the molecular orientation is a reasonable approach  for small tilt angles, since the pre-edge and the near-edge spectra are affected differently by the molecule-surface electronic interactions.

Our XANES results in general show  good agreement the experimental findings of the nitrogen K-edge spectra obtained by  Lovat {\it et al.}\cite{CSLovat2019}. For instance, (i) the absence (presence) of pre-edge absorption feature A$^\prime$ in the upstanding (flat-lying) configurations of  NHC/Au, and (ii) the raise of the N($1s$)\,$\rightarrow$\,$\sigma^\ast$ transition (B$^\prime$ and C$^\prime$) at about  3\,eV above A$^\prime$. These  features are in good agreement with our XANES simulations, being consistent with the projected density of states (PDOS in  Fig.\,\ref{pdos}). However, in (ii) they measured  a  tilt angle $\theta$ of $\sim$40$^\circ$, supported by DFT adsorption energy calculations, while we found the ground state configuration for $\theta=27^\circ$. Here, we can point out two aspects that may explain such a disagreement, one is a possible presence of a multi-phase system in \iPr/Au, composed by \emph{tilted} and \emph{complex} configurations, which can be translated  as larger tilt angles in XANES spectra. In fact, multi-phase systems, made by \emph{tilted} and \emph{complex} geometries of  \iPr\, on the Au(111) surface, have been observed by Inayeh {\it et al.}\cite{CHEMRXIVinayeh2020}. With respect to the DFT results, it is reported  that the adsorption energy weakly depends on the tilt angle\,\cite{CSLovat2019}, and thus, we may have some  uncertainties on the   ground states geometry as a function of $\theta$, since it may depends on the adequacy of the formalism  used to calculate the vdW contribution, and other calculation parameters like the size of the surface unit cell. In addition, as shown in Fig.\,\ref{iPr}(c), there is  azimuthal   dependence ($\phi$) on the total energy which should be taken into account for a precise determination of ground state value of $\theta$.

\section{Conclusions}

Based on {\it ab initio} DFT total energy calculations, combined with simulation of X-ray absorption spectra, we performed a thorough study of the energetic stability and structural arrangement of N-heterocyclic carbenes derived from benzimidazole on the Au(111) surface, {\X}/Au (with X=iPr, Et, and Me). We found that \Et\, and \Me\, molecules are more stable lying parallel to the surface plane, forming \emph{complex} structures on the Au(111) surface. Whereas, for \iPr/Au we found that the \emph{tilted} and \emph{complex} configurations  are somewhat close in energy, where the former configuration is more stable than the latter by about 50\,meV/molecule. These results provide a total energy support to the recent experimental findings, (i) the observation of flat-lying \Et/ and \Me/Au and upstanding \iPr/Au\,\cite{CPCLarrea2017}, while (ii) in Refs.\,\cite{CSLovat2019,CHEMRXIVinayeh2020} the authors identified a dual conformation, namely \emph{tilted}  and \emph{complex}, for \iPr\, molecules on Au(111).

Further investigation, based on the nitrogen K-edge XANES simulations  revealed that the  molecular orientation  on the Au(111) surface is characterized by distinct X-ray absorption fingerprints. For a polarization vector perpendicular to the Au(111) surface, we found that the near-edge spectra of \emph{adatom}-{\X}/Au, upstanding configuration, is given solely  by the N($1s$)\,$\rightarrow$\,$\sigma^\ast$ transition, with no contribution from the $\pi^\ast$ orbitals. The absorption intensity of N($1s$)\,$\rightarrow$\,$\sigma^\ast$ transitions reduces  upon the vertical tilt of the molecule, \emph{tilt}-{\X}/Au; followed by the emergence of an absorption peak at the pre-edge region attributed to the N($1s$)\,$\rightarrow$\,$\pi^\ast$ transition. Finally, we show that such an emergence and increase of the  pre-edge absorption feature, concomitantly with the suppression of the N($1s$)\,$\rightarrow$\,$\sigma^\ast$ (near -edge) transitions indicates the formation of {\X} molecules lying parallel to the Au(111) surface, \emph{complex}-{\X}/Au.  In order to improve the understanding of X-ray absorption results, not only the ones presented in this work, but also those obtained experimentally  for \iPr\, and other NHC molecules on Au(111)\cite{CSLovat2019}; the interpretation of the final spectra was performed in light of the (i) the projection of the electronic density of states, and (ii) a set of XANES simulations of hypothetical molecular \X\, structures.

\begin{acknowledgement}
The authors acknowledge many fruitful discussions with Cathleen M. Crudden, Ishwar Singh, Alex J. Veinot, Alex Inayeh and Ryan R. K. Groome. The authors also acknowledge   financial   support   from   the Brazilian  agencies FAPESP (grant 2019/20857-0), CNPq, and FAPEMIG, and the CENAPAD-SP and Laborat\'{o}rio Nacional de Computa\c{c}\~{a}o Cient\'{i}fica (LNCC-SCAFMat2) for computer time. ABM acknowledges financial support from the Natural Sciences and Engineering Research Council of Canada (NSERC) and the Canadian Foundation for Innovation (CFI). 
\end{acknowledgement}

%
%
%

\bibliography{bib}

\providecommand{\latin}[1]{#1}
\makeatletter
\providecommand{\doi}
  {\begingroup\let\do\@makeother\dospecials
  \catcode`\{=1 \catcode`\}=2 \doi@aux}
\providecommand{\doi@aux}[1]{\endgroup\texttt{#1}}
\makeatother
\providecommand*\mcitethebibliography{\thebibliography}
\csname @ifundefined\endcsname{endmcitethebibliography}
  {\let\endmcitethebibliography\endthebibliography}{}
\begin{mcitethebibliography}{44}
\providecommand*\natexlab[1]{#1}
\providecommand*\mciteSetBstSublistMode[1]{}
\providecommand*\mciteSetBstMaxWidthForm[2]{}
\providecommand*\mciteBstWouldAddEndPuncttrue
  {\def\EndOfBibitem{\unskip.}}
\providecommand*\mciteBstWouldAddEndPunctfalse
  {\let\EndOfBibitem\relax}
\providecommand*\mciteSetBstMidEndSepPunct[3]{}
\providecommand*\mciteSetBstSublistLabelBeginEnd[3]{}
\providecommand*\EndOfBibitem{}
\mciteSetBstSublistMode{f}
\mciteSetBstMaxWidthForm{subitem}{(\alph{mcitesubitemcount})}
\mciteSetBstSublistLabelBeginEnd
  {\mcitemaxwidthsubitemform\space}
  {\relax}
  {\relax}

\bibitem[Weidner \latin{et~al.}(2011)Weidner, Baio, Mundstock, Gro{\ss}e,
  Karth{\"a}user, Bruhn, and Siemeling]{WeidnerAJC2011}
Weidner,~T.; Baio,~J.~E.; Mundstock,~A.; Gro{\ss}e,~C.; Karth{\"a}user,~S.;
  Bruhn,~C.; Siemeling,~U. {NHC-Based Self-Assembled Monolayers on Solid Gold
  Substrates.} \emph{Australian Journal of Chemistry} \textbf{2011}, \emph{64},
  1177--1179\relax
\mciteBstWouldAddEndPuncttrue
\mciteSetBstMidEndSepPunct{\mcitedefaultmidpunct}
{\mcitedefaultendpunct}{\mcitedefaultseppunct}\relax
\EndOfBibitem
\bibitem[Zhukhovitskiy \latin{et~al.}(2013)Zhukhovitskiy, Mavros, Van~Voorhis,
  and Johnson]{ZhukhovitskiyJACS2013}
Zhukhovitskiy,~A.~V.; Mavros,~M.~G.; Van~Voorhis,~T.; Johnson,~J.~A.
  {Addressable Carbene Anchors for Gold Surfaces}. \emph{Journal of the
  American Chemical Society} \textbf{2013}, \emph{135}, 7418--7421\relax
\mciteBstWouldAddEndPuncttrue
\mciteSetBstMidEndSepPunct{\mcitedefaultmidpunct}
{\mcitedefaultendpunct}{\mcitedefaultseppunct}\relax
\EndOfBibitem
\bibitem[Crudden \latin{et~al.}(2014)Crudden, Horton, Ebralidze, Zenkina,
  McLean, Drevniok, She, Kraatz, Mosey, Seki, Keske, Leake, Rousina-Webb, and
  Wu]{NATURECrudden2014}
Crudden,~C.~M.; Horton,~J.~H.; Ebralidze,~I.~I.; Zenkina,~O.~V.; McLean,~A.~B.;
  Drevniok,~B.; She,~Z.; Kraatz,~H.-B.; Mosey,~N.~J.; Seki,~T. \latin{et~al.}
  {Ultra stable self-assembled monolayers of N-heterocyclic carbenes on gold}.
  \emph{Nature Chemistry} \textbf{2014}, \emph{6}, 409\relax
\mciteBstWouldAddEndPuncttrue
\mciteSetBstMidEndSepPunct{\mcitedefaultmidpunct}
{\mcitedefaultendpunct}{\mcitedefaultseppunct}\relax
\EndOfBibitem
\bibitem[Zhukhovitskiy \latin{et~al.}(2015)Zhukhovitskiy, MacLeod, and
  Johnson]{ZhukhovitskiyChemRev2015}
Zhukhovitskiy,~A.~V.; MacLeod,~M.~J.; Johnson,~J.~A. {Carbene Ligands in
  Surface Chemistry: From Stabilization of Discrete Elemental Allotropes to
  Modification of Nanoscale and Bulk Substrates}. \emph{Chem. Rev.}
  \textbf{2015}, \emph{115}, 11503--11532\relax
\mciteBstWouldAddEndPuncttrue
\mciteSetBstMidEndSepPunct{\mcitedefaultmidpunct}
{\mcitedefaultendpunct}{\mcitedefaultseppunct}\relax
\EndOfBibitem
\bibitem[Crudden \latin{et~al.}(2016)Crudden, Horton, Narouz, Li, Smith, Munro,
  Baddeley, Larrea, Drevniok, Thanabalasingam, McLean, Zenkina, Ebralidze, She,
  Kraatz, Mosey, Saunders, and Yagi]{NATURECrudden2016}
Crudden,~C.~M.; Horton,~J.~H.; Narouz,~M.~R.; Li,~Z.; Smith,~C.~A.; Munro,~K.;
  Baddeley,~C.~J.; Larrea,~C.~R.; Drevniok,~B.; Thanabalasingam,~B.
  \latin{et~al.}  {Simple direct formation of self-assembled N-heterocyclic
  carbene monolayers on gold and their application in biosensing}. \emph{Nature
  Communications} \textbf{2016}, \emph{7}, 12654\relax
\mciteBstWouldAddEndPuncttrue
\mciteSetBstMidEndSepPunct{\mcitedefaultmidpunct}
{\mcitedefaultendpunct}{\mcitedefaultseppunct}\relax
\EndOfBibitem
\bibitem[R{\"u}hling \latin{et~al.}(2016)R{\"u}hling, Schaepe, Rakers,
  Vonh{\"o}ren, Tegeder, Ravoo, and Glorius]{RuhlingAC2016}
R{\"u}hling,~A.; Schaepe,~K.; Rakers,~L.; Vonh{\"o}ren,~B.; Tegeder,~P.;
  Ravoo,~B.~J.; Glorius,~F. {Modular Bidentate Hybrid NHC-Thioether Ligands for
  the Stabilization of Palladium Nanoparticles in Various Solvents.}
  \emph{Angewandte Chemie International Edition} \textbf{2016}, \emph{55},
  5856--5860\relax
\mciteBstWouldAddEndPuncttrue
\mciteSetBstMidEndSepPunct{\mcitedefaultmidpunct}
{\mcitedefaultendpunct}{\mcitedefaultseppunct}\relax
\EndOfBibitem
\bibitem[Wang \latin{et~al.}(2016)Wang, R{\"u}hling, Amirjalayer, Knor, Ernst,
  Richter, Gao, Timmer, Gao, Doltsinis, Glorius, and Fuchs]{NATUREWang2016}
Wang,~G.; R{\"u}hling,~A.; Amirjalayer,~S.; Knor,~M.; Ernst,~J.~B.;
  Richter,~C.; Gao,~H.-J.; Timmer,~A.; Gao,~H.-Y.; Doltsinis,~N.~L.
  \latin{et~al.}  {Ballbot-type motion of N-heterocyclic carbenes on gold
  surfaces}. \emph{Nature Chemistry} \textbf{2016}, \emph{9}, 152\relax
\mciteBstWouldAddEndPuncttrue
\mciteSetBstMidEndSepPunct{\mcitedefaultmidpunct}
{\mcitedefaultendpunct}{\mcitedefaultseppunct}\relax
\EndOfBibitem
\bibitem[Engel \latin{et~al.}(2017)Engel, Fritz, and
  Ravoo]{EngelChemSocRev2017}
Engel,~S.; Fritz,~E.-C.; Ravoo,~B.~J. {New trends in the functionalization of
  metallic gold: from organosulfur ligands to N-heterocyclic carbenes.}
  \emph{Chemical Society Reviews} \textbf{2017}, \emph{46}, 2057--2075\relax
\mciteBstWouldAddEndPuncttrue
\mciteSetBstMidEndSepPunct{\mcitedefaultmidpunct}
{\mcitedefaultendpunct}{\mcitedefaultseppunct}\relax
\EndOfBibitem
\bibitem[Salorinne \latin{et~al.}(2017)Salorinne, Man, Li, Taki, Nambo, and
  Crudden]{SalorinneAC2017}
Salorinne,~K.; Man,~R. W.~Y.; Li,~C.-H.; Taki,~M.; Nambo,~M.; Crudden,~C.~M.
  {Water-Soluble N-Heterocyclic Carbene-Protected Gold Nanoparticles:
  Size-Controlled Synthesis, Stability, and Optical Properties.}
  \emph{Angewandte Chemie International Edition} \textbf{2017}, \emph{56},
  6198--6202\relax
\mciteBstWouldAddEndPuncttrue
\mciteSetBstMidEndSepPunct{\mcitedefaultmidpunct}
{\mcitedefaultendpunct}{\mcitedefaultseppunct}\relax
\EndOfBibitem
\bibitem[Larrea \latin{et~al.}(2017)Larrea, Baddeley, Narouz, Mosey, Horton,
  and Crudden]{CPCLarrea2017}
Larrea,~C.~R.; Baddeley,~C.~J.; Narouz,~M.~R.; Mosey,~N.~J.; Horton,~J.~H.;
  Crudden,~C.~M. {N-Heterocyclic Carbene Self-assembled Monolayers on Copper
  and Gold: Dramatic Effect of Wingtip Groups on Binding, Orientation and
  Assembly}. \emph{ChemPhysChem} \textbf{2017}, \emph{18}, 3536--3539\relax
\mciteBstWouldAddEndPuncttrue
\mciteSetBstMidEndSepPunct{\mcitedefaultmidpunct}
{\mcitedefaultendpunct}{\mcitedefaultseppunct}\relax
\EndOfBibitem
\bibitem[Man \latin{et~al.}(2018)Man, Li, MacLean, Zenkina, Zamora, Saunders,
  Rousina-Webb, Nambo, and Crudden]{ManJACS2018}
Man,~R. W.~Y.; Li,~C.-H.; MacLean,~M. W.~A.; Zenkina,~O.~V.; Zamora,~M.~T.;
  Saunders,~L.~N.; Rousina-Webb,~A.; Nambo,~M.; Crudden,~C.~M. {Ultrastable
  Gold Nanoparticles Modified by Bidentate N-Heterocyclic Carbene Ligands}.
  \emph{Journal of the American Chemical Society} \textbf{2018}, \emph{140},
  1576--1579\relax
\mciteBstWouldAddEndPuncttrue
\mciteSetBstMidEndSepPunct{\mcitedefaultmidpunct}
{\mcitedefaultendpunct}{\mcitedefaultseppunct}\relax
\EndOfBibitem
\bibitem[{H{\"a}kkinen, Hannu}(2012)]{HakkinenNature2012}
{H{\"a}kkinen, Hannu}, The gold-sulfur interface at the nanoscale. \emph{Nature
  Chemistry} \textbf{2012}, \emph{4}, 443--455\relax
\mciteBstWouldAddEndPuncttrue
\mciteSetBstMidEndSepPunct{\mcitedefaultmidpunct}
{\mcitedefaultendpunct}{\mcitedefaultseppunct}\relax
\EndOfBibitem
\bibitem[Casalini \latin{et~al.}(2017)Casalini, Bortolotti, Leonardi, and
  Biscarini]{CSRcasalini2017}
Casalini,~S.; Bortolotti,~C.~A.; Leonardi,~F.; Biscarini,~F. {Self-assembled
  monolayers in organic electronics}. \emph{Chem. Soc. Rev.} \textbf{2017},
  \emph{46}, 40--71\relax
\mciteBstWouldAddEndPuncttrue
\mciteSetBstMidEndSepPunct{\mcitedefaultmidpunct}
{\mcitedefaultendpunct}{\mcitedefaultseppunct}\relax
\EndOfBibitem
\bibitem[Mandler and Kraus-Ophir(2011)Mandler, and
  Kraus-Ophir]{JSSEMandler2011}
Mandler,~D.; Kraus-Ophir,~S. {Self-assembled monolayers (SAMs) for
  electrochemical sensing}. \emph{Journal of Solid State Electrochemistry}
  \textbf{2011}, \emph{15}, 1535\relax
\mciteBstWouldAddEndPuncttrue
\mciteSetBstMidEndSepPunct{\mcitedefaultmidpunct}
{\mcitedefaultendpunct}{\mcitedefaultseppunct}\relax
\EndOfBibitem
\bibitem[Schoenbaum \latin{et~al.}(2014)Schoenbaum, Schwartz, and
  Medlin]{ACRschoenbaum2014}
Schoenbaum,~C.~A.; Schwartz,~D.~K.; Medlin,~J.~W. {Controlling the Surface
  Environment of Heterogeneous Catalysts Using Self-Assembled Monolayers}.
  \emph{Accounts of Chemical Research} \textbf{2014}, \emph{47},
  1438--1445\relax
\mciteBstWouldAddEndPuncttrue
\mciteSetBstMidEndSepPunct{\mcitedefaultmidpunct}
{\mcitedefaultendpunct}{\mcitedefaultseppunct}\relax
\EndOfBibitem
\bibitem[Smith \latin{et~al.}(2019)Smith, Narouz, Lummis, Singh, Nazemi, Li,
  and Crudden]{CRSmith2019}
Smith,~C.~A.; Narouz,~M.~R.; Lummis,~P.~A.; Singh,~I.; Nazemi,~A.; Li,~C.-H.;
  Crudden,~C.~M. {N-Heterocyclic Carbenes in Materials Chemistry}.
  \emph{Chemical Reviews} \textbf{2019}, \emph{119}, 4986--5056\relax
\mciteBstWouldAddEndPuncttrue
\mciteSetBstMidEndSepPunct{\mcitedefaultmidpunct}
{\mcitedefaultendpunct}{\mcitedefaultseppunct}\relax
\EndOfBibitem
\bibitem[Liang \latin{et~al.}(2019)Liang, shun qi, Zhang, A, Chen, Cui, Qi,
  Sun, and Zhao]{NATCOMMliang2019}
Liang,~R.-R.; shun qi,~x.; Zhang,~L.; A,~R.-H.; Chen,~P.; Cui,~F.-Z.;
  Qi,~Q.-Y.; Sun,~J.; Zhao,~X. {Rational design of crystalline two-dimensional
  frameworks with highly complicated topological structures}. \emph{Nature
  Communications} \textbf{2019}, \emph{10}, 1--9\relax
\mciteBstWouldAddEndPuncttrue
\mciteSetBstMidEndSepPunct{\mcitedefaultmidpunct}
{\mcitedefaultendpunct}{\mcitedefaultseppunct}\relax
\EndOfBibitem
\bibitem[Bakker \latin{et~al.}(2018)Bakker, Timmer, Kolodzeiski, Freitag, Gao,
  M\"onig, Amirjalayer, Glorius, and Fuchs]{wangNatChem2017}
Bakker,~A.; Timmer,~A.; Kolodzeiski,~E.; Freitag,~M.; Gao,~H.~Y.; M\"onig,~H.;
  Amirjalayer,~S.; Glorius,~F.; Fuchs,~H. {Elucidating the Binding Modes of
  N-Heterocyclic Carbenes on a Gold Surface}. \emph{Journal of the American
  Chemical Society} \textbf{2018}, \emph{140}, 11889\relax
\mciteBstWouldAddEndPuncttrue
\mciteSetBstMidEndSepPunct{\mcitedefaultmidpunct}
{\mcitedefaultendpunct}{\mcitedefaultseppunct}\relax
\EndOfBibitem
\bibitem[Lovat \latin{et~al.}(2019)Lovat, Doud, Lu, Kladnik, Inkpen,
  Steigerwald, Cvetko, Hybertsen, Morgante, Roy, and Venkataraman]{CSLovat2019}
Lovat,~G.; Doud,~E.~A.; Lu,~D.; Kladnik,~G.; Inkpen,~M.~S.; Steigerwald,~M.~L.;
  Cvetko,~D.; Hybertsen,~M.~S.; Morgante,~A.; Roy,~X. \latin{et~al.}
  {Determination of the structure and geometry of N-heterocyclic carbenes on
  Au(111) using high-resolution spectroscopy}. \emph{Chem. Sci.} \textbf{2019},
  \emph{10}, 930--935\relax
\mciteBstWouldAddEndPuncttrue
\mciteSetBstMidEndSepPunct{\mcitedefaultmidpunct}
{\mcitedefaultendpunct}{\mcitedefaultseppunct}\relax
\EndOfBibitem
\bibitem[Inayeh \latin{et~al.}(2020)Inayeh, Groome, Singh, Veinot, Lima, Miwa,
  Crudden, and McLean]{CHEMRXIVinayeh2020}
Inayeh,~A.; Groome,~R.; Singh,~I.; Veinot,~A.; Lima,~F.; Miwa,~R.; Crudden,~C.;
  McLean,~A. {Self-Assembly of N-Heterocyclic Carbenes on Au(111)}.
  \emph{ChemRxiv} \textbf{2020}, \relax
\mciteBstWouldAddEndPunctfalse
\mciteSetBstMidEndSepPunct{\mcitedefaultmidpunct}
{}{\mcitedefaultseppunct}\relax
\EndOfBibitem
\bibitem[Jiang \latin{et~al.}(2017)Jiang, Zhang, M\'edard, Seitsonen, Haag,
  Allegretti, Reichert, Kuster, Barth, and Papageorgiou]{jiangChemSci2017}
Jiang,~L.; Zhang,~B.; M\'edard,~G.; Seitsonen,~A.~P.; Haag,~F.; Allegretti,~F.;
  Reichert,~J.; Kuster,~B.; Barth,~J.~V.; Papageorgiou,~A.~C. {N-Heterocyclic
  carbenes on close-packed coinage metal surfaces: bis-carbene metal adatom
  bonding scheme of monolayer films on Au{,} Ag and Cu}. \emph{Chem. Sci.}
  \textbf{2017}, \emph{8}, 8301\relax
\mciteBstWouldAddEndPuncttrue
\mciteSetBstMidEndSepPunct{\mcitedefaultmidpunct}
{\mcitedefaultendpunct}{\mcitedefaultseppunct}\relax
\EndOfBibitem
\bibitem[Krzykawska \latin{et~al.}(2020)Krzykawska, Wr{\'o}bel, Kozie{\l}, and
  Cyganik]{krzykawskaACSNano2020}
Krzykawska,~A.; Wr{\'o}bel,~M.; Kozie{\l},~K.; Cyganik,~P. N-Heterocyclic
  Carbenes for the Self-Assembly of Thin and Highly Insulating Monolayers with
  High Quality and Stability. \emph{ACS nano} \textbf{2020}, \emph{14},
  6043--6057\relax
\mciteBstWouldAddEndPuncttrue
\mciteSetBstMidEndSepPunct{\mcitedefaultmidpunct}
{\mcitedefaultendpunct}{\mcitedefaultseppunct}\relax
\EndOfBibitem
\bibitem[Kresse and Furthm{\"u}ller(1996)Kresse, and Furthm{\"u}ller]{VASP}
Kresse,~G.; Furthm{\"u}ller,~J. {Efficiency of ab-initio total energy
  calculations for metals and semiconductors using a plane-wave basis set}.
  \emph{Computational Materials Science} \textbf{1996}, \emph{6}, 15 --
  50\relax
\mciteBstWouldAddEndPuncttrue
\mciteSetBstMidEndSepPunct{\mcitedefaultmidpunct}
{\mcitedefaultendpunct}{\mcitedefaultseppunct}\relax
\EndOfBibitem
\bibitem[Perdew \latin{et~al.}(1996)Perdew, Burke, and Ernzerhof]{PBE}
Perdew,~J.~P.; Burke,~K.; Ernzerhof,~M. {Generalized Gradient Approximation
  Made Simple}. \emph{Phys. Rev. Lett.} \textbf{1996}, \emph{77}, 3865\relax
\mciteBstWouldAddEndPuncttrue
\mciteSetBstMidEndSepPunct{\mcitedefaultmidpunct}
{\mcitedefaultendpunct}{\mcitedefaultseppunct}\relax
\EndOfBibitem
\bibitem[Monkhorst and Pack(1976)Monkhorst, and Pack]{PhysRevB.13.5188}
Monkhorst,~H.~J.; Pack,~J.~D. {Special points for Brillouin-zone integrations}.
  \emph{Phys. Rev. B} \textbf{1976}, \emph{13}, 5188--5192\relax
\mciteBstWouldAddEndPuncttrue
\mciteSetBstMidEndSepPunct{\mcitedefaultmidpunct}
{\mcitedefaultendpunct}{\mcitedefaultseppunct}\relax
\EndOfBibitem
\bibitem[Bl\"ochl(1994)]{PAW}
Bl\"ochl,~P.~E. {Projector augmented-wave method}. \emph{Phys. Rev. B}
  \textbf{1994}, \emph{50}, 17953--17979\relax
\mciteBstWouldAddEndPuncttrue
\mciteSetBstMidEndSepPunct{\mcitedefaultmidpunct}
{\mcitedefaultendpunct}{\mcitedefaultseppunct}\relax
\EndOfBibitem
\bibitem[Dion \latin{et~al.}(2004)Dion, Rydberg, Schr\"oder, Langreth, and
  Lundqvist]{PhysRevLett.92.246401}
Dion,~M.; Rydberg,~H.; Schr\"oder,~E.; Langreth,~D.~C.; Lundqvist,~B.~I. {Van
  der Waals Density Functional for General Geometries}. \emph{Phys. Rev. Lett.}
  \textbf{2004}, \emph{92}, 246401\relax
\mciteBstWouldAddEndPuncttrue
\mciteSetBstMidEndSepPunct{\mcitedefaultmidpunct}
{\mcitedefaultendpunct}{\mcitedefaultseppunct}\relax
\EndOfBibitem
\bibitem[Klime{\v{s}} \latin{et~al.}(2009)Klime{\v{s}}, Bowler, and
  Michaelides]{klimesJPhysC2010}
Klime{\v{s}},~J.; Bowler,~D.~R.; Michaelides,~A. {Chemical accuracy for the van
  der Waals density functional}. \emph{Journal of Physics: Condensed Matter}
  \textbf{2009}, \emph{22}, 022201\relax
\mciteBstWouldAddEndPuncttrue
\mciteSetBstMidEndSepPunct{\mcitedefaultmidpunct}
{\mcitedefaultendpunct}{\mcitedefaultseppunct}\relax
\EndOfBibitem
\bibitem[Carrasco \latin{et~al.}(2014)Carrasco, Liu, Michaelides, and
  Tkatchenko]{JCPcarrasco2014}
Carrasco,~J.; Liu,~W.; Michaelides,~A.; Tkatchenko,~A. {Insight into the
  description of van der Waals forces for benzene adsorption on transition
  metal (111) surfaces}. \emph{The Journal of Chemical Physics} \textbf{2014},
  \emph{140}, 084704\relax
\mciteBstWouldAddEndPuncttrue
\mciteSetBstMidEndSepPunct{\mcitedefaultmidpunct}
{\mcitedefaultendpunct}{\mcitedefaultseppunct}\relax
\EndOfBibitem
\bibitem[Lee \latin{et~al.}(2010)Lee, Murray, Kong, Lundqvist, and
  Langreth]{VDW-DF2}
Lee,~K.; Murray,~E.~D.; Kong,~L.; Lundqvist,~B.~I.; Langreth,~D.~C.
  {Higher-accuracy van der Waals density functional}. \emph{Phys. Rev. B}
  \textbf{2010}, \emph{82}, 081101\relax
\mciteBstWouldAddEndPuncttrue
\mciteSetBstMidEndSepPunct{\mcitedefaultmidpunct}
{\mcitedefaultendpunct}{\mcitedefaultseppunct}\relax
\EndOfBibitem
\bibitem[Taillefumier \latin{et~al.}(2002)Taillefumier, Cabaret, Flank, and
  Mauri]{PhysRevB.66.195107}
Taillefumier,~M.; Cabaret,~D.; Flank,~A.-M.; Mauri,~F. {X-ray absorption
  near-edge structure calculations with the pseudopotentials: Application to
  the \textit{K} edge in diamond and $\ensuremath{\alpha}$-quartz}. \emph{Phys.
  Rev. B} \textbf{2002}, \emph{66}, 195107\relax
\mciteBstWouldAddEndPuncttrue
\mciteSetBstMidEndSepPunct{\mcitedefaultmidpunct}
{\mcitedefaultendpunct}{\mcitedefaultseppunct}\relax
\EndOfBibitem
\bibitem[Giannozzi \latin{et~al.}(2009)Giannozzi, \latin{et~al.} others]{qe}
others,, \latin{et~al.}  {QUANTUM ESPRESSO: a modular and open-source software
  project for quantum simulations of materials}. \emph{Journal of Physics:
  Condensed Matter} \textbf{2009}, \emph{21}, 395502\relax
\mciteBstWouldAddEndPuncttrue
\mciteSetBstMidEndSepPunct{\mcitedefaultmidpunct}
{\mcitedefaultendpunct}{\mcitedefaultseppunct}\relax
\EndOfBibitem
\bibitem[Pickard and Mauri(2001)Pickard, and Mauri]{PRBpickard2001}
Pickard,~C.~J.; Mauri,~F. All-electron magnetic response with pseudopotentials:
  NMR chemical shifts. \emph{Phys. Rev. B} \textbf{2001}, \emph{63},
  245101\relax
\mciteBstWouldAddEndPuncttrue
\mciteSetBstMidEndSepPunct{\mcitedefaultmidpunct}
{\mcitedefaultendpunct}{\mcitedefaultseppunct}\relax
\EndOfBibitem
\bibitem[Brouder(1990)]{JPCMbrouder1990}
Brouder,~C. Angular dependence of X-ray absorption spectra. \emph{Journal of
  Physics: Condensed Matter} \textbf{1990}, \emph{2}, 701--738\relax
\mciteBstWouldAddEndPuncttrue
\mciteSetBstMidEndSepPunct{\mcitedefaultmidpunct}
{\mcitedefaultendpunct}{\mcitedefaultseppunct}\relax
\EndOfBibitem
\bibitem[Taillefumier \latin{et~al.}(2002)Taillefumier, Cabaret, Flank, and
  Mauri]{PRBtaillefumier2002}
Taillefumier,~M.; Cabaret,~D.; Flank,~A.-M.; Mauri,~F. X-ray absorption
  near-edge structure calculations with the pseudopotentials: Application to
  the K edge in diamond and $\ensuremath{\alpha}$-quartz. \emph{Phys. Rev. B}
  \textbf{2002}, \emph{66}, 195107\relax
\mciteBstWouldAddEndPuncttrue
\mciteSetBstMidEndSepPunct{\mcitedefaultmidpunct}
{\mcitedefaultendpunct}{\mcitedefaultseppunct}\relax
\EndOfBibitem
\bibitem[Gougoussis \latin{et~al.}(2009)Gougoussis, Calandra, Seitsonen, and
  Mauri]{PRBgougoussis2009}
Gougoussis,~C.; Calandra,~M.; Seitsonen,~A.~P.; Mauri,~F. {First-principles
  calculations of x-ray absorption in a scheme based on ultrasoft
  pseudopotentials: From $\ensuremath{\alpha}$-quartz to high-${T}_{c}$
  compounds}. \emph{Phys. Rev. B} \textbf{2009}, \emph{80}, 075102\relax
\mciteBstWouldAddEndPuncttrue
\mciteSetBstMidEndSepPunct{\mcitedefaultmidpunct}
{\mcitedefaultendpunct}{\mcitedefaultseppunct}\relax
\EndOfBibitem
\bibitem[Rodr{\'{i}}guez-Castillo \latin{et~al.}(2016)Rodr{\'{i}}guez-Castillo,
  Lugo-Preciado, Laurencin, Tielens, van~der Lee, Cl{\'{e}}ment, Guari,
  de~Luzuriaga, Monge, Remacle, and Richeter]{CAEJRodriguez2016}
Rodr{\'{i}}guez-Castillo,~M.; Lugo-Preciado,~G.; Laurencin,~D.; Tielens,~F.;
  van~der Lee,~A.; Cl{\'{e}}ment,~S.; Guari,~Y.; de~Luzuriaga,~J. M.~L.;
  Monge,~M.; Remacle,~F. \latin{et~al.}  {Experimental and Theoretical Study of
  the Reactivity of Gold Nanoparticles Towards Benzimidazole-2-ylidene
  Ligands}. \emph{Chemistry - A European Journal} \textbf{2016}, \emph{22},
  10446--10458\relax
\mciteBstWouldAddEndPuncttrue
\mciteSetBstMidEndSepPunct{\mcitedefaultmidpunct}
{\mcitedefaultendpunct}{\mcitedefaultseppunct}\relax
\EndOfBibitem
\bibitem[Tang and Jiang(2017)Tang, and Jiang]{CMTang2017}
Tang,~Q.; Jiang,~D.-e. {Comprehensive View of the Ligand–Gold Interface from
  First Principles}. \emph{Chemistry of Materials} \textbf{2017}, \emph{29},
  6908--6915\relax
\mciteBstWouldAddEndPuncttrue
\mciteSetBstMidEndSepPunct{\mcitedefaultmidpunct}
{\mcitedefaultendpunct}{\mcitedefaultseppunct}\relax
\EndOfBibitem
\bibitem[Grimme(2006)]{grimme2006semiempirical}
Grimme,~S. Semiempirical GGA-type density functional constructed with a
  long-range dispersion correction. \emph{Journal of computational chemistry}
  \textbf{2006}, \emph{27}, 1787--1799\relax
\mciteBstWouldAddEndPuncttrue
\mciteSetBstMidEndSepPunct{\mcitedefaultmidpunct}
{\mcitedefaultendpunct}{\mcitedefaultseppunct}\relax
\EndOfBibitem
\bibitem[Grimme \latin{et~al.}(2010)Grimme, Antony, Ehrlich, and
  Krieg]{grimme2010consistent}
Grimme,~S.; Antony,~J.; Ehrlich,~S.; Krieg,~H. A consistent and accurate ab
  initio parametrization of density functional dispersion correction (DFT-D)
  for the 94 elements H-Pu. \emph{The Journal of chemical physics}
  \textbf{2010}, \emph{132}, 154104\relax
\mciteBstWouldAddEndPuncttrue
\mciteSetBstMidEndSepPunct{\mcitedefaultmidpunct}
{\mcitedefaultendpunct}{\mcitedefaultseppunct}\relax
\EndOfBibitem
\bibitem[Bakker \latin{et~al.}(2018)Bakker, Timmer, Kolodzeiski, Freitag, Gao,
  Mönig, Amirjalayer, Glorius, and Fuchs]{JACSBakker2018}
Bakker,~A.; Timmer,~A.; Kolodzeiski,~E.; Freitag,~M.; Gao,~H.~Y.; Mönig,~H.;
  Amirjalayer,~S.; Glorius,~F.; Fuchs,~H. {Elucidating the Binding Modes of
  N-Heterocyclic Carbenes on a Gold Surface}. \emph{Journal of the American
  Chemical Society} \textbf{2018}, \emph{140}, 11889--11892\relax
\mciteBstWouldAddEndPuncttrue
\mciteSetBstMidEndSepPunct{\mcitedefaultmidpunct}
{\mcitedefaultendpunct}{\mcitedefaultseppunct}\relax
\EndOfBibitem
\bibitem[Bakker \latin{et~al.}(2018)Bakker, Timmer, Kolodzeiski, Freitag, Gao,
  M\"onig, Amirjalayer, Glorius, and Fuchs]{bakkerJACS2018}
Bakker,~A.; Timmer,~A.; Kolodzeiski,~E.; Freitag,~M.; Gao,~H.~Y.; M\"onig,~H.;
  Amirjalayer,~S.; Glorius,~F.; Fuchs,~H. {Elucidating the Binding Modes of
  N-Heterocyclic Carbenes on a Gold Surface}. \emph{Journal of the American
  Chemical Society} \textbf{2018}, \emph{140}, 11889\relax
\mciteBstWouldAddEndPuncttrue
\mciteSetBstMidEndSepPunct{\mcitedefaultmidpunct}
{\mcitedefaultendpunct}{\mcitedefaultseppunct}\relax
\EndOfBibitem
\bibitem[Chang \latin{et~al.}(2017)Chang, Chen, Lu, and Cheng]{changJPCA2017}
Chang,~K.; Chen,~J.~G.; Lu,~Q.; Cheng,~M.-J. Quantum Mechanical Study of
  N-Heterocyclic Carbene Adsorption on Au Surfaces. \emph{The Journal of
  Physical Chemistry A} \textbf{2017}, \emph{121}, 2674--2682\relax
\mciteBstWouldAddEndPuncttrue
\mciteSetBstMidEndSepPunct{\mcitedefaultmidpunct}
{\mcitedefaultendpunct}{\mcitedefaultseppunct}\relax
\EndOfBibitem
\end{mcitethebibliography}

\end{document}